\documentstyle[aps]{revtex}

\begin{document}

\title{Transport mean free path for Magneto-Transverse Light Diffusion: an alternative approach}
\author{David Lacoste \\
Department of Physics, University of Pennsylvania, Philadelphia, PA 19104}

 \maketitle

\begin{abstract}
 This article presents a derivation of the transport mean 
 free path  for  magneto-transverse light diffusion,
 $\ell^*_\perp$,
in an arbitrary random mixture of Faraday-active
and non-Faraday active Mie scatterers. 
This derivation is based on the standard radiative transfer equation.
The expression of the transport mean free path obtained previously
 from the Bethe-Salpether equation,
for the case where only Faraday-active scatterers are present,
is recovered. This simpler formulation can include the
 case of homogeneous mixtures of Faraday-active and 
non-Faraday active scatterers.

\end{abstract}

\section{Introduction}
Magneto-transverse light diffusion - also
known as the ``Photonic Hall Effect" (PHE) - was theoretically predicted
five years ago by Van Tiggelen \cite{bart4}, 
and was experimentally confirmed 
one year later by Rikken \cite{nature}. 
This effect is 
analogous to the electronic Hall effect, well known in semiconductors physics.
The evident driver behind the electronic Hall effect
is the Lorentz force acting on charged particles.
 The PHE finds its origin in the Faraday effect, present inside the dielectric
 scatterers, which slightly changes their scattering amplitude. 
This is the reason for the similarities between the PHE and the 
so-called Beenakker-Senftleben effect, which concerns
the transport coefficients
 of dilute paramagnetic gazes \cite{beenakker}.

Motivated by the experimental observation of the PHE, 
theoretical work
was first started on the single scattering of
spherical magneto-optical particles.
The scattering matrix and the scattering cross-section were
calculated exactly for a single Faraday-active dielectric sphere
  using perturbation theory \cite{josa}. From this solution, 
the Stokes parameters, which completely describe the intensity and
polarization of the scattered light, were derived \cite{jqsr}.
Perturbational methods for the scattering by weakly anisotropic particles 
were first used by Kuzmin {\it et al.} for the single scattering case \cite{kuzmin}.

The PHE in multiple scattering 
is controlled by a length, which has been defined as
the mean free path for magneto-transverse light diffusion, $\ell^*_\perp$.
A simple expression for this length was obtained and successfully
compared
 to experiments using a formulation
 based on the ladder approximation of the Bethe-Salpether 
equation \cite{euro}. The experiments investigated the
dependence of the PHE on the volume fraction of the scatterers,
 first 
on the real part of the dielectric constant of the scatterers \cite{nature},
 and more recently on their imaginary part \cite{sabine2}.
However, some parts of the derivation of Ref. \cite{euro} are rather
 technical and fail to give a satisfying explanation of the origin of the 
dependence of $\ell^*_\perp$
 on the anisotropy of the scattering.
This article presents a simpler derivation of $\ell^*_\perp$,
based on the radiative transfer equation, which should clarify this point.
 The radiative transfer equation is a Boltzmann type equation which
describes the transport of light in multiple light scattering \cite{chandrasekhar,hulst2}.
From the radiative transfer equation, the diffusion equation
is derived in an infinite medium when the scattering is not highly 
anisotropic.
In other types of anisotropic materials, such as single-domain nematic
liquid crystals, similar methods were developed, using 
either the Bethe-Salpether equation \cite{heiderich} or
the radiative transfer equation \cite{holger}.

\section{Single scattering}
The single scattering of light by one dielectric sphere
made of a Faraday-active material embedded in an isotropic
medium with no magneto-optical properties is first considered.
In a magnetic field, the dielectric constant of the sphere, $\epsilon_B$, is a
tensor of rank two. It depends on the distance to the center
of the sphere of radius, $R$, via the Heaviside function,
$ \Theta (|{\mathbf{r}}|-R)$, which equals $1$ inside the sphere and
$0$ outside \cite{josa},
\begin{equation}
\epsilon_B({\mathbf{B}},{\mathbf{r}})_{ij}=\left[
(\epsilon_0-1)\delta_{ij}+i\epsilon_F \epsilon_{ijk} 
 \hat{B_k} \right] \Theta (|{\mathbf{r}}|-R),
\end{equation}
where $\epsilon_0$ is the value of the normal isotropic
dielectric constant of the sphere, 
$\epsilon_F=2{\epsilon_0}^{1/2} V_0 B/\omega$
is the coupling constant of the Faraday effect,
$\delta$ and $\epsilon$ are respectively the Kronecker and Levi-Civitta tensors.
 The Verdet constant 
of the Faraday effect is denoted by $V_0$, $B$ is the amplitude of the magnetic field
 and $\omega$ is the frequency.
The intensity of the scattered light, 
in single scattering, is characterized by the phase function which is also the
essential ingredient for the transport theory of light.
The phase function, $F(\hat{\mathbf{k}},\hat{\mathbf{k}}'
,{\mathbf{B}})$, is proportional to the differential scattering
cross-section averaged with respect to incoming and outgoing
polarizations.  It depends on the direction of the incoming plane wave,
$\widehat{\mathbf{k}}$, on the direction of the scattered wave,
$\widehat{\mathbf{k}}'$, and on the direction of the magnetic field,
${\mathbf{B}}$. The hat above the vectors denotes normalized vectors.
For spherical magneto-optical particles and to linear order in the
applied magnetic field, this phase function can be written as
\cite{josa}
\begin{equation}
\label{fctphase:mag}
F(\hat{\mathbf{k}},\hat{\mathbf{k}}',{\mathbf{B}})=F_{0}(\hat{\mathbf{k}},
\hat{\mathbf{k}}')+\mathrm{det}(\hat{\mathbf{k}},\hat{\mathbf{k}}',
\widehat{{\mathbf{B}}})\, F_{1}(\hat{\mathbf{k}},\hat{\mathbf{k}}')\, ,
\end{equation}
where $\mathrm{det}(\mathbf{A}, 
 {{\mathbf{B}}}, {\mathbf{C}})=  {\mathbf{A}}
\cdot ( {{\mathbf{B}}} \times  {\mathbf{C}})$ denotes the scalar
determinant. 
Due to the rotational symmetry of the scatterer, the functions, $F_0$
and $F_{1}$, only depend on the scattering angle, $\theta$, which is
the angle between $\widehat{\mathbf{k}}$ and $\widehat{\mathbf{k}}'$.
The phase function $F_1$ only depends on the difference
in the azimuthal angles associated with
$\widehat{\mathbf{k}}$ and $\widehat{\mathbf{k}}'$,
because of the axial symmetry of the scatterer around the direction of
the magnetic field. These symmetry properties 
simplify considerably the use of the radiative transfer equation
\cite{mishchenko}.  The amplitude of $F_1$ was found to be
proportional to the dimensionless parameter, $\epsilon_F$ \cite{josa}.

The albedo is introduced as 
the ratio of the scattering cross-section over the extinction
cross-section \cite{hulst}
\begin{equation}
a={Q_{scatt} \over Q_{ext}}.
\end{equation}
It is related to the phase function
 defined in Eq.~(\ref{fctphase:mag}), 
\begin{equation}
\label{albedo}
a=\int d\widehat{\mathbf{k}} \,
 F(\widehat{\mathbf{k}}, 
\widehat{\mathbf{k}}', \widehat{{\mathbf{B}}}) =
\int d\widehat{\mathbf{k}} \,
 F_0(\widehat{\mathbf{k}}, 
\widehat{\mathbf{k}}'),
\end{equation}
where the last equality follows from Eq.~(\ref{fctphase:mag}).
Thus in the absence of absorption the phase function
is normalized $(a=1)$.

\section{Transport in magneto-optical diffuse media}
In an isotropic medium much larger than the transport mean free path,
 $\ell^*$, Fick's law relates the diffusion current, ${\mathbf{J}}$, to 
the energy-density gradient, $\nabla I$, by 
${\mathbf{J}} = -D_0 \nabla I $. 
The conventional diffusion constant
for radiative transfer equation is denoted
$D_0$ and is usually 
related to the transport mean free path, $\ell^*$, and 
the transport velocity, $v_E$, by $D_0= \frac13 v_E \ell^*$. 
In the presence of a magnetic field,
the diffusion constant is a second-rank tensor.
The part of this tensor linear in the magnetic field 
is responsible for the PHE, whereas the
 part quadratic in the magnetic field 
generates a photonic magneto-resistance, which could also be observed
experimentally \cite{anja}.
By Onsager's relation, $D_{ij}({\mathbf{B}}) =D_{ji}(\mathbf{-B})$, the part linear in the external magnetic field
must be an  antisymmetric tensor, and Fick's law becomes,
	\begin{equation}
	{\mathbf{J}} = - {\mathbf{D}}({\mathbf{B}}) \cdot \nabla I =  
	-D_0\nabla I - D_\perp \hat{{\mathbf{B}}} \times \nabla I\, .\label{symm}
	\end{equation}
The    term containing $D_\perp$ 
describes the magneto-transverse diffusion current responsible for the PHE.
 In analogy to the definition of $\ell^*$,
 the transport mean free path, $\ell^*_\perp$, for magneto-transverse
 diffusion
is defined as $D_\perp= \frac13 v_E \ell^*_\perp$.
  For the electronic Hall effect,  $\ell^*_\perp$
is proportional  to the Hall conductivity,  $\sigma_{xy}$, 
the sign of which
is determined by the charge of the current carriers.  
Similarly,   a positive $\ell^*_\perp$ means that the PHE 
has the same sign as the Verdet constant of the scatterers, and a negative
 $\ell^*_\perp$ means that the PHE has an opposite sign, which
is also possible depending on the scattering.
This point has been carefully checked experimentally \cite{nature}.

The phase function introduced above can be used to describe not
 only the single scattering of light by independently scattering
 particles but also light
 scattering by a large collection of such particles
in the regime of multiple light scattering. 
It is assumed that only magneto-optical Mie scatterers are present,
and included in a matrix with no magneto-optical properties.
A transport equation for a medium comprising 
 randomly distributed spherical
particles embedded in an isotropic medium can be written
 as follows \cite{rossum}:
\begin{equation}
\label{TR:magn}
\frac{\ell}{v_E} \, \partial _{t}{\mathcal{I}}({\mathbf{r}},
\hat{\mathbf{k}},t)+\ell \, 
\hat{\mathbf{k}}\cdot \nabla {\mathcal{I}}({\mathbf{r}},\hat{\mathbf{k}},
t)+{\mathcal{I}}({\mathbf{r}},\hat{\mathbf{k}},t)=
\int d\hat{\mathbf{k}}'\,
 F(\hat{\mathbf{k}},\hat{\mathbf{k}}',{\mathbf{B}}){\mathcal{I}}({\mathbf{r}},\hat{\mathbf{k}}',
t),
\end{equation}
where $\ell$ denotes the elastic mean free path, which
 is the average distance between two subsequent scattering events.
The specific intensity 
${\mathcal{I}}({\mathbf{r}},\hat{\mathbf{k}},t)$ 
is defined as the density of radiation at position
 ${\mathbf{r}}$, time $t$, in the direction 
$\hat{\mathbf{k}}$. The specific intensity 
has the dimension of a monochromatic radiance:
energy per unit solid angle, time, wavelength and surface area \cite{travis}.
When supplemented by appropriate boundary conditions, the radiative transfer
equation can be solved for a particular problem.
It is useful to introduce the average specific intensity,
$I({\mathbf{r}},t)$, and the current, ${\mathbf{J}}({\mathbf{r}},t)$,
which are defined as 

\begin{equation}
\label{defJ}
I({\mathbf{r}},t)=\int {\mathcal{I}}({\mathbf{r}},
\hat{\mathbf{k}},t)d\hat{\mathbf{k}},\, \, \, \,
 \, \, \, \, \, \, \, \, \, \, \, \, {\mathbf{J}}
({\mathbf{r}},t)=v_{E}\, \int \hat{\mathbf{k}}d\hat{\mathbf{k}}\,
 {\mathcal{I}}({\mathbf{r}},\hat{\mathbf{k}},t).
\end{equation}
To simplify the notations, the dependence of the
specific intensity or of related quantities upon the
 magnetic field, ${\mathbf{B}}$, is not
explicitly indicated.
The integration of Eq.~(\ref{TR:magn})
with respect to $\hat{\mathbf{k}}$
leads to the continuity equation
\begin{equation}
\label{continuity}
\partial _{t}I({\mathbf{r}},t)+\nabla \cdot {\mathbf{J}}({\mathbf{r}},t)
=-\frac{(1-a)v_{E}}{\ell }I({\mathbf{r}},t),
\end{equation}
where $a$ denotes the albedo defined in Eq.~(\ref{albedo}).

If Eq.~(\ref{TR:magn}) is multiplied by
$\hat{\mathbf{k}}$ 
and integrated over the direction
$\hat{\mathbf{k}}$,
the right-hand side of Eq.~(\ref{TR:magn}) will contain two integrals
which depend on 
$F_{0}$ and $F_{1}$.
 These integrals are the two dimensionless quantities:
\begin{equation}
<\cos \theta >=\int d\hat{\mathbf{k}}F_{0}(\hat{\mathbf{k}},\hat{\mathbf{k}}')\, 
\hat{\mathbf{k}}\cdot \hat{\mathbf{k}}',
\end{equation}
and
\begin{equation}
\label{equ:A1}
A_1=\int d\hat{\mathbf{k}}F_{1}(\hat{\mathbf{k}},\hat{\mathbf{k}}')\,
 \mathrm{det}(\hat{\mathbf{k}},\hat{\mathbf{k}}',\widehat{{\mathbf{B}}})^{2}.
\end{equation}
By choosing a system of coordinates linked with 
$\hat{\mathbf{k}}'$, it can be proved quite generally that
\begin{equation}
\int d\hat{\mathbf{k}}F(\hat{\mathbf{k}},\hat{\mathbf{k}}',\hat{{\mathbf{B}}})\,
 \hat{\mathbf{k}}=<\cos \theta >
\hat{\mathbf{k}}'-A_{1}\hat{{\mathbf{B}}}
\times \hat{\mathbf{k}}'.
\end{equation}
After the integration over 
$\hat{\mathbf{k}}'$ the following equation is obtained
	\begin{equation}
	\label{eq:courant}
	\frac{\ell}{v_E} \, \partial _{t}
	{\mathbf{J}}+
	{\mathbf{J}}\, \left( 1-<\cos \theta >\right)
	 +A_{1}\hat{{\mathbf{B}}}\times {\mathbf{J}}=-\ell v_{E}\,
	 \nabla \cdot \int d\hat{\mathbf{k}}\,
	 {\mathcal{I}}({\mathbf{r}},\hat{\mathbf{k}},t)\hat{\mathbf{k}}\hat{\mathbf{k}}.
	\end{equation}
If it is assumed that the angular distribution
 of the specific intensity is almost isotropic, the current is much
smaller than the average specific intensity.
In that case the following approximation is valid:
\cite{ishimaru} 
	\begin{equation}
	\label{Tr:dev:magn}
	{\mathcal{I}}({\mathbf{r}},\hat{\mathbf{k}},t)\approx
	 I({\mathbf{r}},t)+\frac{3}{v_{E}}{\mathbf{J}}({\mathbf{r}},
	t)\cdot \hat{\mathbf{k}}+....
	\end{equation} 
Substituting this relation into Eq.~(\ref{eq:courant}), and 
neglecting $\partial_t {\mathbf{J}}$ for processes
 which are slow with respect to the characteristic time between two 
scatterings, $\ell /v_{E}$,
gives Fick's law \cite{rossum}:
\begin{equation}
\label{Fick}
{\mathbf{J}}({\mathbf{r}},t)=-{\mathbf{D}} \cdot \nabla I({\mathbf{r}},t).
\end{equation} 
The diffusion tensor is given by,
\begin{equation}
\label{DdeB}
D({\mathbf{B}})_{ij}=\frac{1}{3}v_{E}\ell \, \left[ \left( 1-<\cos \theta >\right)
 \delta_{ij}-A_{1}\varepsilon_{ijk} \hat{B_k}\right] ^{-1}.
\end{equation}
It is important to note that the magnetic correction to the scattering cross-section
plays a role similar to the role of 
the asymmetry factor, $<\cos \theta >$, present when no
magnetic field is applied.
The dependence of  $<\cos \theta >$ can be obtained from
standard Mie theory \cite{hulst}.
Both anisotropy factors, $<\cos \theta >$ and $A_1$, 
are non-zero provided that the symmetry between forward and backward
scattering is broken, which is the case for a finite size scatterer.
This is the reason why the PHE of a Rayleigh
scatterer vanishes whereas the PHE of a Mie scatterer
does not \cite{josa}.

 In the case where 
$A_{1}\ll (1-<\cos \theta >)$,
 an expansion of the bracket gives
\begin{equation}
\label{Diff:magn:scalaire}
D_{ij}({\mathbf{B}})=\frac{1}{3}v_{E}\frac{\ell }{1-<\cos \theta >}\delta
 _{ij}+\frac{1}{3}v_{E}A_{1}\frac{\ell }{\left( 1-<\cos \theta >\right) ^{2}}\varepsilon _{ijk}\hat{B}_{k}.
\end{equation} 
This equation identifies 
$\ell ^{*}=\ell /(1-<\cos \theta >)$ with the transport mean free path, 
and 
$\ell _{\perp } ^{*}=A_{1}\ell ^{*}/(1-<\cos \theta >)$
 as the transport mean free path for magneto-transverse diffusion.
The final result of Ref. \cite{euro} is thus recovered with little efforts.
The order of magnitude of the ratio $\ell _{\perp } ^{*}/\ell ^{*}$
is $10^{-5}$ for an applied magnetic field of $1$T and for 
samples made of rare-earth materials
 \cite{nature}.
 The presence in Eq.~(\ref{Diff:magn:scalaire}) of the coefficient, $A_{1}$, 
which is directly connected with the PHE of a single scatterer \cite{josa},
 states that the PHE in multiple scattering 
is directly proportional
to the normalized PHE of a single Mie sphere, 
is of the same sign and amplified by the factor,
$1/(1-<\cos \theta >)^{2}$.
The expansion to first order in the magnetic field 
involved in Eq.~(\ref{Diff:magn:scalaire}),
clarifies the origin of the
factor, $1/(1-<\cos \theta >)^{2}$, which makes
 $\ell _{\perp }^{*}$ 
even more dependent on the asymmetry factor,
$<\cos \theta >$ 
than the transport mean free path,
$\ell ^{*}$.

Substituting Eq.~(\ref{Fick}) into the continuity equation 
(\ref{continuity})
yields the diffusion equation for the density of radiation
\begin{equation}
\label{eq:diffusion}
\partial _{t}I({\mathbf{r}},t)=D_0 \nabla ^{2}I({\mathbf{r}},t)-
\frac{(1-a)v_{E}}{\ell }I({\mathbf{r}},t)=
D_0 \left[ \nabla ^{2}
I({\mathbf{r}},t)-\frac{1}{L_{a}^{2}}I({\mathbf{r}},t) \right].
\end{equation}
As expected, the part of the diffusion tensor, which is linear
in the magnetic field and therefore antisymmetric does not enter
 the diffusion equation which depends only on the symmetric part, $D_0$.
In addition to the transport mean free path, $\ell^{*}$, 
Eq.~(\ref{eq:diffusion}) defines the characteristic length for the absorption in
multiple light scattering, $L_a$
\begin{equation}
\label{def:La}
L_{a}=\sqrt{\frac{\ell ^{*}\cdot \ell }{3(1-a)}}=\sqrt{\frac{\ell ^{*}\cdot \ell _{abs}}{3}},
\end{equation}
as function of the characteristic length for the absorption of 
the coherent beam  
$\ell _{abs}$.
Therefore there is no dependence of $L_a$ on the magnetic field to linear order.
As discussed in Ref.~\cite{furutsu}, the diffusion coefficient $D_0$ does not depend 
on the absorption cross-section, which is proportional to $1-a$. 
Eq.~(\ref{Diff:magn:scalaire}) shows that 
this property also holds for $\ell_{\perp}^{*}$ when the medium is dissipative.

\section{Transport in mixtures  of
Faraday-active and non Faraday-active spheres}
The main result of Eq.~(\ref{Diff:magn:scalaire}) can be
extended to the case of an homogeneous mixture of
Faraday-active spheres with volume fraction, $n_0$, 
and non Faraday-active spheres
with volume fraction, $n_1$. 
The phase function of Eq.~(\ref{fctphase:mag}) is now
modified for the case of the mixtures as follows
\begin{eqnarray}
\label{fctphase:mel}
F(\hat{\mathbf{k}},\hat{\mathbf{k}}',\hat{{\mathbf{B}}})_{mix} & = &
{n_0 \over {n_0 + n_1}}
F_{0}(\hat{\mathbf{k}},
\hat{\mathbf{k}}')+
{n_1 \over {n_0+n_1}}
\left[
F_{0}(\hat{\mathbf{k}},
\hat{\mathbf{k}}') +
\mathrm{det}(\hat{\mathbf{k}},\hat{\mathbf{k}}',
\widehat{{\mathbf{B}}})\, 
F_{1}(\hat{\mathbf{k}},\hat{\mathbf{k}}')\, 
\right] \nonumber, \\
& = &
F_{0}(\hat{\mathbf{k}},
\hat{\mathbf{k}}')+
{n_1 \over {n_0+n_1}}
\mathrm{det}(\hat{\mathbf{k}},\hat{\mathbf{k}}',
\widehat{{\mathbf{B}}})\, 
F_{1}(\hat{\mathbf{k}},\hat{\mathbf{k}}').
\end{eqnarray}
The equation (\ref{DdeB}) can be generalized to
\begin{equation}
\label{Dgen}
D({\mathbf{B}})_{ij}=\frac{1}{3}v_{E}\ell \, \left[
(1-<\cos \theta >) \delta_{ij}
-{n_1 \over {n_0+n_1}}
A_{1}\varepsilon_{ijk} \hat{B_k}\right] ^{-1}.
\end{equation}
In the case when the last term of the bracket is small with respect 
to $1-<\cos \theta >$,
the following expression for the ratio of the transport
mean free path for magneto-transverse diffusion over the transport mean free path
 is obtained
\begin{equation}
{\ell_{\perp }^{*} \over \ell^{*}}=
{n_1 \over {n_0+n_1}}\,
{A_1 \over { 1-<\cos \theta > }}.
\end{equation}

\section{Conclusion}
In conclusion, the present transport 
theory for magneto-optical spheres of arbitrary size
 confirms the results of Ref.~\cite{euro}
concerning the magneto-transverse light diffusion
obtained from the Bethe-Salpether equation.
This new approach clarifies the origin of the dependence of
the transport mean free path for magneto-transverse diffusion
on the anisotropy of the scattering.
It also shows that the magnetic anisotropy in the
scattering cross-section plays a similar role as 
 $\left<\cos\theta\right>$ in the transport mean free path.
This derivation can be generalized to the case of mixtures
of Faraday-active and non Faraday-active scatterers.

\section{Acknowledgments}
I thank  B. Van Tiggelen, G. Rikken, S. Wiebel and J-J. Greffet for
the many stimulating discussions. This work has been made 
possible by the Groupement de Recherches POAN.  
  

\end{document}